\documentclass[%
 reprint,
 superscriptaddress,
 amsmath,amssymb,
 aps,
 floatfix,
]{revtex4-2}

\usepackage{dcolumn}% Align table columns on decimal point
\usepackage{bm}% bold math
\usepackage{here}
   \usepackage[pdftex]{graphicx}

\begin{document}

%\preprint{APS/123-QED}

\title{Radiation-tolerant polarized solid target}% Force line breaks with \\

\author{K.~Tateishi}
\altaffiliation[kenichiro.tateishi@riken.jp]{}
\affiliation{Nishina Center for Accelerator-Based Science, RIKEN, Saitama 351-0198, Japan}
\affiliation{Pioneering Research Institute, RIKEN, Saitama 351-0198, Japan}
%\email{kenichiro.tateishi@riken.jp}

\author{Y.~Saito}
%\affiliation{Department of Physics, Tohoku University, Miyagi 980-8578, Japan}
\affiliation{Nishina Center for Accelerator-Based Science, RIKEN, Saitama 351-0198, Japan}

\author{D.~Takahashi}
\affiliation{Department of Physics, Institute of Science Tokyo, Tokyo 152-8551, Japan}

\author{K.~Sekiguchi}
\affiliation{Nishina Center for Accelerator-Based Science, RIKEN, Saitama 351-0198, Japan}
%\affiliation{Department of Physics, Tohoku University, Miyagi 980-8578, Japan}
\affiliation{Department of Physics, Institute of Science Tokyo, Tokyo 152-8551, Japan}
\affiliation{Department of Physics, Kyoto University, Kyoto 606-8502, Japan}

\author{K.~Aradono}
\affiliation{Department of Physics, Kyushu University, Fukuoka 819-0395, Japan}

\author{K.~Hirasawa}
\affiliation{Department of Physics, Kyushu University, Fukuoka 819-0395, Japan}

\author{Y.~Maeda}
\affiliation{Department of Applied Physics, University of Miyazaki, Miyazaki, 889-2192, Japan}

\author{Y.~Nagao}
\affiliation{Department of Physics, Kyushu University, Fukuoka 819-0395, Japan}

\author{H.~Nishibata}
\affiliation{Department of Physics, Kyushu University, Fukuoka 819-0395, Japan}

\author{S.~Otsuka}
\affiliation{Pioneering Research Institute, RIKEN, Saitama 351-0198, Japan}
\affiliation{Department of Physics, Saitama University, Saitama 338-8570, Japan}

\author{H.~Sakai}
\affiliation{Nishina Center for Accelerator-Based Science, RIKEN, Saitama 351-0198, Japan}

%\author{N.~Sakamoto}
%\affiliation{Nishina Center for Accelerator-Based Science, RIKEN, Saitama 351-0198, Japan}

\author{H.~Sugahara}
\affiliation{Department of Physics, Institute of Science Tokyo, Tokyo 152-8551, Japan}

\author{K.~Suzuki}
\affiliation{Department of Physics, Institute of Science Tokyo, Tokyo 152-8551, Japan}

\author{T.~Uesaka} 
\affiliation{Nishina Center for Accelerator-Based Science, RIKEN, Saitama 351-0198, Japan}
\affiliation{Pioneering Research Institute, RIKEN, Saitama 351-0198, Japan}
\affiliation{Department of Physics, Saitama University, Saitama 338-8570, Japan}

\author{T.~Wakasa}
\affiliation{Department of Physics, Kyushu University, Fukuoka 819-0395, Japan}

\author{A.~Watanabe}
%\affiliation{Department of Physics, Institute of Science Tokyo, Tokyo 152-8551, Japan}
\affiliation{Nishina Center for Accelerator-Based Science, RIKEN, Saitama 351-0198, Japan}

\date{\today}% It is always \today, today,

\begin{abstract}
    Polarized targets evolved into indispensable tools in particle and nuclear physics.
    However, the polarized solid target is degraded by high-intense beam irradiation, known as radiation damage due to target heating and radical generation.
    We demonstrated a radiation-tolerant polarized solid target operating at room temperature. 
    An annealing allows the spontaneous repair of the damage by reducing unwanted radicals.
    Using a single crystal of $\it p$-terphenyl doped with 0.01 mol\% pentacene-$\it d$$_{14}$, Dynamic Nuclear Polarization using photoexcited triplet electrons (Triplet-DNP) was applied to proton spins at room temperature and in 0.39 T.
    For the proof of concept, a deuteron beam with an energy of 135 MeV/u and the intensities of 10$^7$-10$^9$ counts per second (cps) was irradiated.
    The proton polarization was determined to be 3.0\% $\pm$0.2\%$\rm{{(stat.)}}$ $\pm$0.1\%$\rm {{(sys.)}}$ from a scattering asymmetry.
    The polarization was almost not attenuated up to 10$^9$ cps, but the target crystal was yellowed.
    The visible-light absorption spectroscopy suggested irreversible radiation damage due to missing protons by the knock-out reaction.
    The room-temperature polarized solid target allows impractical experiments with the conventional target system, leading to a next-generation spin-dependent accelerator science.
\end{abstract}

\maketitle

%\tableofcontents

Within decades, polarized solid targets as well as polarized beams evolved into indispensable tools to explore the structure and reaction in particle and nuclear physics. 
Recently, the origin of proton spin has been intensively investigated from the viewpoint of quantum chromodynamics (QCD) using them~\cite{Andrieux}. 
The polarization also gives us important information about spin-dependent interactions, such as spin-orbit, tensor, and three-body forces in nuclei~\cite{Uesaka,Watanabe}. 
However, the small magnetic moment of nuclei leads to significant disturbances from thermal energy, resulting in only slight polarization. 
Thus, several methods for dynamic polarization of nuclei have been developed~\cite{Eills}.

Dynamic nuclear polarization (DNP), a means of transferring spin polarization from electrons to nuclei by microwave irradiation, is applied to polarized solid targets~\cite{Abragam,Crabb}. 
The solid target is high density, and has less background, but requires high-tech equipment compared to the gas target.
The target material has to be cooled to cryogenic temperatures below 1 K and placed in a magnetic field of several Tesla ($>$2.5 T) to polarize thermally-equilibrated electrons~\cite{Andrieux}. 
Proton spins can polarize almost unity with DNP under these conditions but are degraded by high-intense beam irradiation, called radiation damage. 
The damage originates from two effects, target heating~\cite{Thomas, Liu} and radical generation~\cite{Andrieux, Meyer}. 
The former became more serious at cryogenic temperatures because of smaller heat capacity. 
To suppress the latter, a variety of materials were developed such as lanthanum magnesium nitrate (LMN), butanol, ammonia, and lithium hydride~\cite{Goertz}. 
The collision with the high-energy beam particles induces chemical bond breaking of the target material, producing radicals that accelerate spin relaxation. 
%Consequently, the radiation damage strongly limits the beam intensity, $\it {i.e.}$ statistical accuracy of the experiment.
Consequently, the radiation damage strongly limits the beam intensity, and thus the statistical accuracy of the experiment.

To solve the problem, we proposed the radiation-tolerant polarized solid target. 
The idea was spontaneous repair of the damage by ``annealing''. 
The reduction of the unwanted radicals in ammonia was known by heating the target to 75-95 K~\cite{McKee}.
The mechanism was carefully investigated and confirmed to occur around 190 K in organic materials~\cite{Capozzi}, indicating the radiation damage can be $\it {effectively}$ canceled if the repairing rate exceeds the damage.  

\begin{figure}[t]
  \centering
  \includegraphics[width=1\columnwidth]{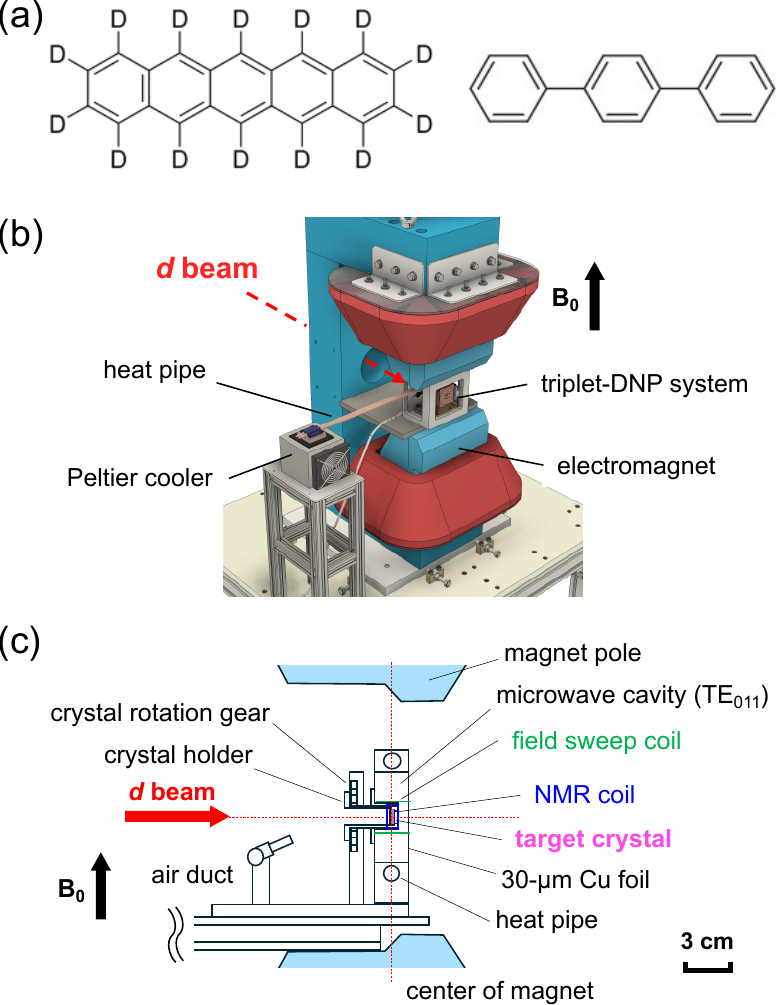}
  \caption{
   (a) Molecular structures of pentacene-$\it d$$_{14}$ (C$_{22}$D$_{14}$) and $\it p$-terphenyl (C$_{18}$H$_{14}$). 
   (b) Polarized proton solid target based on Triplet-DNP operatering at room temterature and in 0.39 T.
   (c) Side view of Triplet-DNP system installed inside the electromagnet.
   }
  %\label{fig_3}
\end{figure}

In this paper, we present the polarized solid target with DNP using photoexcited triplet electrons (Triplet-DNP) operating at room temperature ~\cite{Henstra,Takeda,Hamachi}. 
The photoexcited triplet electrons have large non-equilibrated polarization independent of temperature and magnetic field, $\it {e.g.}$ $>$70\% for pentacene-$\it d$$_{14}$ which is the standard polarizing agent of Triplet-DNP~\cite{Ong}. 
This method has been applied to the neutron spin filter~\cite{Quan,Hautle,Takada} and the polarized target for light-ion beams~\cite{Uesaka,Watanabe24}. 
Proton spins were polarized up to 80\% at 25 K and in 0.3 T using a single crystal of naphthalene doped with pentacene-$\it d$$_{14}$~\cite{Quan}. 
That of 30\% was obtained even at 100 K~\cite{Watanabe24}, however, the naphthalene was rapidly evaporated at room temperature by laser irradiation. 
Therefore, we introduced $\it p$-terphenyl as a host molecule of the target~\cite{Tateishi}. 
The new system operates at room temperature and 0.39 T. 
For the proof of concept, a 135 MeV/u deuteron beam was irradiated with the intensities of 10$^7$-10$^9$ counts per second (cps). 
The NMR signal, which is proportional to the polarization, was kept constant during the beam irradiation, as expected.

The procedure of Triplet-DNP in 0.39 T was as follows.
%Details were previously reported in Ref. [].
Pentacene-$\it d$$_{14}$ was photoexcited with a 556 nm pulsed laser.
Then, a 10.2 GHz microwave was irradiated to transfer the polarization from electron spins to proton spins in $\it p$-terphenyl. 
The external field was swept during the microwave irradiation to improve the transfer efficiency. 
The sign of proton polarization can be varied by switching the sweep phase.
This scheme was called the Integrated Solid Effect (ISE)~\cite{Henstra14}. 
By repeating the ISE sequence, the proton polarization can be buildup until it is balanced with the spin relaxation.
A single crystal of $\it p$-terphenyl doped with 0.01 mol\% pentacene-$\it d$$_{14}$ was grown with the Bridgman method (Fig.~1(a)). 
The crystal with a size of $\phi$10 $\times$ 2.5 mm was mounted in a Teflon holder and set the long axis of the pentacene-$\it d$$_{14}$ parallel to the external magnetic field.
The Triplet-DNP system is shown in Fig.~1(b)(c). 
The electromagnet was designed for the detection of both longitudinally and transversely scattered particles. 
A diode-pumped solid-state laser (CNI, HPL-589-Q) was used for pentacene-$\it d$$_{14}$ excitation. 
The wavelength, pulse width, pulse energy, and repetition rate were 556 nm, $\sim$600 ns, 4.0 mJ, and 1.0 kHz, respectively. 
The resonance frequencies of triplet electron and proton in 0.39~T were 10.2 GHz and 16.7 MHz, respectively.
A 20-$\mu$s microwave pulse amplified to 2.2 kW using a pulsed traveling wave tube amplifier (IFI, PT188-1KW) was applied, while the field was adiabatically swept with a voltage of ±50V.
The polarized proton signals were monitored using an OPENCORE nuclear magnetic resonance (NMR) spectrometer~\cite{TakedaOC}. 
The technical difficulties were to prevent the target crystal and cavity heating due to high-power laser and microwave irradiation.
The high temperature accelerates a spin relaxation, resulting in decreasing the proton polarization.
For the crystal, air was blown with a rate of $\sim$200 L/min.
For the cavity, a Peltier cooler (TAISEI, UT-7070J-HS) was utilized via heat pipes.

%A deuteron beam with an energy of 135 MeV/u was irradiated at RIKEN Accelerator Research Facility (RARF) to confirm the spontaneous repair of the radiation damage and to determine the proton polarization from scattering asymmetry.
%Details of the experimental setup are described in the Supporting Information.
%The deuteron beam was transported to the E3A course in the E3 experimental hall.
%Plastic scintillators for the beam counter and beam monitor were placed upstream of the target to monitor the relative beam rates during the experiment. 
%Plastic scintillators for the scattered deuterons and recoil protons in kinematical coincidence condition were positioned downstream at $\theta_{\rm{c.m.}}= 68.8^{\circ}$, $78.6^{\circ}$, $88.6^{\circ}$, $98.6^{\circ}$, $108.8^{\circ}$, $119.0^{\circ}$, $129.4^{\circ}$, $139.8^{\circ}$, and $150.3^{\circ}$ in the center-of-mass system with the azimuthal angles of $\phi = 0^{\circ}$, $90^{\circ}$, $180^{\circ}$, and $270^{\circ}$, respectively.
%Multi-wire drift chambers (MWDCs) were also installed before the downstream scintillators to subtract the background events.
%The beam stopped using a tungsten-based beam dump, placed 2.5 m downstream of the target.

A deuteron beam with an energy of 135 MeV/u was irradiated at RIKEN RI Beam Factory (RIBF) to confirm the spontaneous repair of the radiation damage of the polarized target and to determine the proton polarization from scattering asymmetry. The deuteron beam was transported to the E3A course in the E3 experimental room.
The polarimetry was made using deuteron-proton elastic scattering~\cite{Ay1,Ay2}.
%Plastic scintillators for the beam counter were placed upstream of the target to monitor the relative beam rates during the experiment. 
The scattered deuterons and recoil protons were detected under kinematical coincidence conditions using a detector system consisting of plastic scintillators and multi-wire drift chambers, arranged symmetrically in the left, right, up, and down directions relative to the beam direction~\cite{Yuko}.
The measured angles were $68.8^{\circ}$, $78.6^{\circ}$, $88.6^{\circ}$, $98.6^{\circ}$, $108.8^{\circ}$, $119.0^{\circ}$, $129.4^{\circ}$, $139.8^{\circ}$, and $150.3^{\circ}$ in the center-of-mass system.
The beam was stopped using a tungsten-based beam dump located 2.5 m downstream of the target.
The beam counting rate was extracted by using the existing differential cross-section data for the deuteron-proton elastic scattering~\cite{Sekiguchi2005}. 
Together with this, the relative value of the beam counting rate was monitored by using the plastic scintillator placed upstream of the target.

The 135 MeV/u deuteron beam was irradiated for 25.5 hours at $2.0 \times 10^7$ count per second (cps), 16.0 hours at $7.0 \times 10^7$ cps and 1.0 hours at $1.0 \times 10^9$ cps, respectively.
Figure 2 shows the NMR signal intensities as a function of time, which is proportional to the polarization. 
The signals were taken every 3 minutes.
The target polarization was switched several times between positive, negative, and unpolarized to reduce systematic errors of the proton polarization estimation.
The negative polarization was slightly lower than the positive due to the improper crystal cutting. 
There is almost no signal attenuation for $\it p$-terphenyl target operating at room temperature, whereas a 10\% signal intensity was attenuated in 8 hours at $4 \times 10^6$ cps for the naphthalene target operating at 100 K and in 0.3~T~\cite{Watanabe24}.
The slight attenuation with increasing beam intensity was concluded to be laser power degradation due to the radiational environment.
The laser power decreased from 4.0 to 3.4 mJ after the experiment.
Details will be discussed in later paragraphs.
Thus, a radiation-tolerant polarized solid target has been confirmed.

\begin{figure}[t]
  \centering
  \includegraphics[width=1\columnwidth]{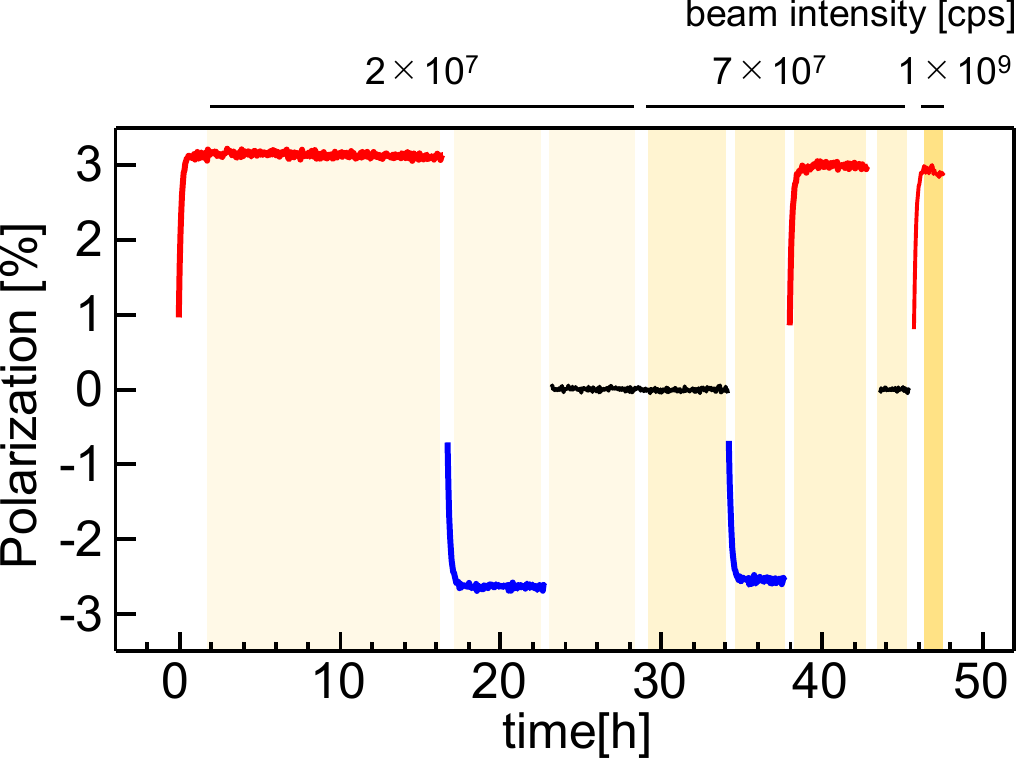}
  \caption{
    Proton polarization buildup curve of the single crystal of $\it p$-terphenyl doped with 0.01 mol\% pentacene-$\it d$$_{14}$ at room temperature and in 0.39 T. 
    %The signals were taken every 3 minutes.
    The polarization was switched several times between positive, negative and unpolarized to reduce systematic errors of the proton polarization estimation.
    The 135 MeV/u deuteron beam was irradiated in the colored areas for 25.5 hours at $2.0 \times 10^7$ cps, 16.0 hours at $7.0 \times 10^7$ cps and 1.0 hours at $1.0 \times 10^9$ cps, respectively.
  }
  \label{fig_buildup}
\end{figure}

The target polarization can be determined from the deuteron-proton scattering asymmetry, independent of the NMR measurements.
The target polarization was determined to be 3.0\% $\pm$0.2\%$\rm{{(stat.)}}$ $\pm$0.1\%$\rm {{(sys.)}}$ averaged over the first ($2.0 \times 10^7$~cps) and second ($7.0 \times 10^7$~cps) irradiations.
The second and third values represent statistic and systematic errors, respectively. 
The polarization was 3.3\% $\pm$0.4\% $\pm$0.1\% in the first weak irradiation, and 2.4\% $\pm$0.3\% $\pm$0.1\% in the second middle, respectively.
Since the detectors could be damaged, the scattered particles were not measured during the third ($1.0 \times 10^9$~cps) strong irradiation.
The fact that the polarization from each measurement agrees within the error indicates that the repairing rate of the unwanted radicals is sufficiently fast, and no local depolarization has occurred.
However, the polarization was not enough high, therefore, updating the target crystal with a longer relaxation time at room temperature~\cite{Tateishi} is an important research direction for future work.

\begin{figure}[t]
  \begin{center}
  \includegraphics[width=1\columnwidth]{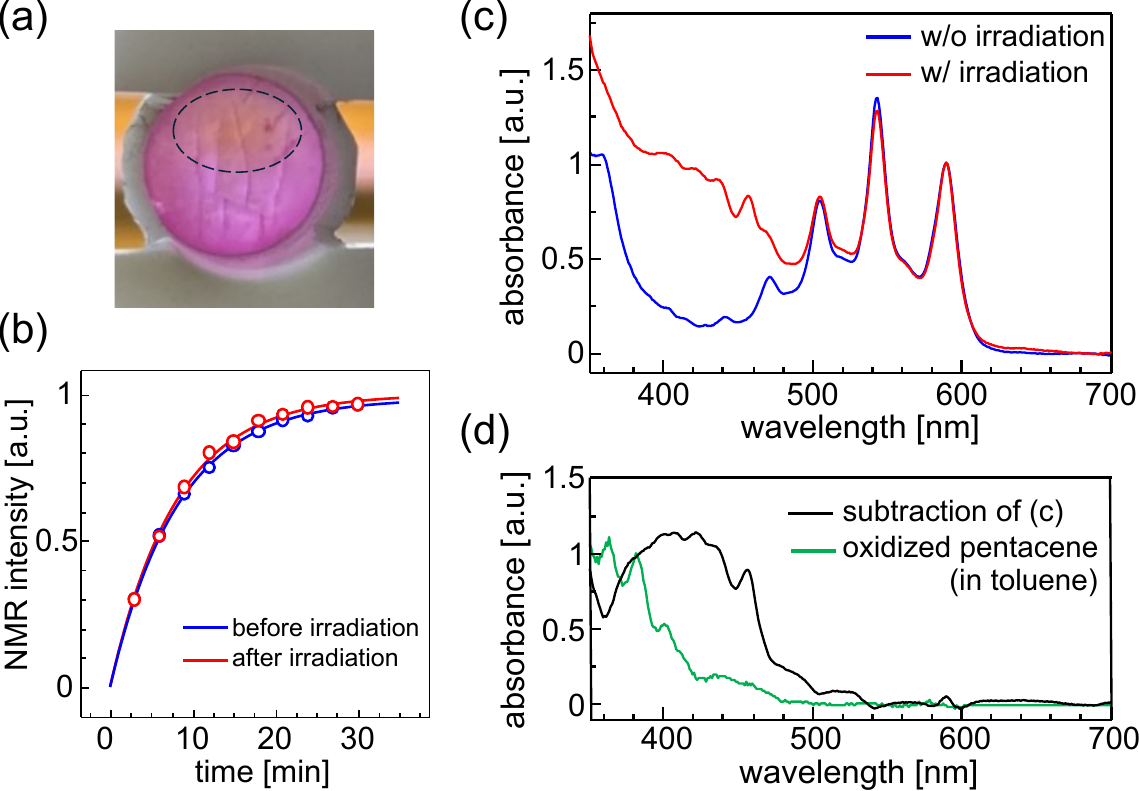}
  \end{center}
  \caption{
   (a) Photo of the target crystal after deuteron beam irradiation.
   (b) Proton polarization buildup curve before (blue) and after (red) deuteron beam irradiation. 
   (c) VIS specra of the target crystal with (red) and without (blue) deuteron beam irradiation.
   (d) VIS spectra of the subtraction of (c) (black) and oxidized pentacene in toluene (green).
  }
  \label{fig_3}
\end{figure}

For a better understanding, the depolarization at high-intense beam was investigated.
The polarization of the third irradiation area was $\sim$5\% lower than the first. 
As mentioned above, the laser power decreased from 4.0 to 3.4 mJ due to the radiational environment.
Consequently, the polarizing power of Triplet-DNP would be reduced.
In addition, the target crystal partially yellowed (Fig.~3(a)).
To clarify that, Triplet-DNP was carried out with the laser power of 4.0 mJ after the experiment.
Figure 3(b) shows the proton polarization buildup curves before and after the beam irradiation.
The initial buildup rate, the buildup constant as well as the finally attainable polarization were reproduced.
Therefore, we concluded that the depolarization at a high-intense beam resulted in laser degradation due to the radiational environment.

We continued the analysis of the yellowing of the target crystal.
There are two possibilities: oxidation of pentacene-$\it d$$_{14}$ by target heating and radical generation in $\it p$-terphenyl with knockout reaction.
Visible-light absorption spectroscopy (VIS) was conducted to address that.
Figure 3(c) shows the VIS spectra of the target crystal with and without the beam irradiation.
The sharp peaks above 440 nm originate from the absorption of pentacene-$\it d$$_{14}$.
$\it p$-Terphenyl dominantly absorbs ultraviolet (UV) light.
The absorption around 350-500 nm increased with the beam irradiation, corresponding to the yellowing of the crystal.
Compared with the spectrum of their subtraction and an oxidized pentacene in toluene, they were much different (Fig.~3(d)).
On the other hand, the subtraction was similar to the spectrum of chemically-generated radicals in aromatic molecules~\cite{Cataldo}.
Consequently, the yellowing is most likely to be radicals in $\it p$-terphenyl.
One possibility for the radicals is due to the missing protons by the knock-out reaction, resulting in irreversible damage.
They induce the depolarization as paramagnetic relaxation sources, but their contribution was not observed in this experiment owing to the small amount.
This will become more serious with increasing beam accumulation.

Finally, this target can be operated not only at high temperatures but also in low magnetic fields, indicating it is suitable for use in nuclear physics experiments, $\it {i.e.}$ low-energy charged particles.
It is more useful for high-intense beams, such as primary beams.
For example, three-body nucleon force research combined with a polarized deuteron beam is one of the promising applications.
The force is essential to clarify nuclear phenomena such as discrete states of nuclei~\cite{HEBELER20211} and the equation of state of the neutron star~\cite{RevModPhys.87.1067}.
Moreover, as large accelerators are constructed around the world, the radiation damage of conventional polarized targets will soon become a bottleneck for statistical accuracy even in secondary beams.
On the other hand, the target size limits the laser intensity, therefore, applications requiring large targets are more difficult.

In summary, we demonstrated the radiation-tolerant polarized solid target operating at room temperature. 
An “annealing” allows the spontaneous repair of the damage by reducing the unwanted radicals.
Triplet-DNP is suitable for the concept because the photoexcited triplet electrons have large non-equilibrated polarization independent of temperature and magnetic field.
We developed the Triplet-DNP based polarized target operating in 0.39 T and at room temperature using a single crystal of $\it p$-terphenyl doped with 0.01 mol\% pentacene-$\it d$$_{14}$.
A deuteron beam with an energy of 135 MeV/u and intensities of 10$^7$-10$^9$ cps were irradiated at RIKEN RIBF.
The proton polarization was determined to be 3.0\% $\pm$0.2\%$\rm{{(stat.)}}$ $\pm$0.1\%$\rm {{(sys.)}}$ from a scattering asymmetry and confirmed the spontaneous repair of radiation damage.
After the experiment, VIS spectra suggested irreversible radiation damage due to missing protons, indicating more serious with increasing beam accumulation.
The room-temperature polarized solid target benefits not only the radiation damage but also the simplification of the experimental setup.
This allows impractical experiments with the conventional target system, leading to a next-generation spin-dependent accelerator science.

We thank the Advanced Manufacturing Support Team at RIKEN for fabricating the microwave resonator.
We also thank the staff of the RIKEN Nishina Center for operating the accelerator.
This work was supported by JSPS KAKENHI (JP20H05636), and JST ERATO Grant No. JPMJER2304, Japan.

\end{document}